\documentclass[aps,prl,reprint,showpacs,showkeys,footinbib,superscriptaddress]{revtex4-1}
\usepackage{amsmath,amssymb,bm}
\usepackage{graphicx}
\usepackage{times,txfonts}
\usepackage{xcolor}
\usepackage{epstopdf}
\usepackage[colorlinks=true,citecolor=blue,bookmarks=false]{hyperref}
\usepackage{lipsum}
\usepackage[all]{hypcap}
\usepackage{CJK}

\begin{document}
\begin{CJK*}{UTF8}{bsmi}

\title{Magnetotransport Signatures of the Radial Rashba Spin-Orbit Coupling in Proximitized Graphene}

\author{Wun-Hao Kang (康文豪)}

\affiliation{Department of Physics, National Cheng Kung University, Tainan 70101, Taiwan}

\affiliation{Center of Quantum Frontiers of Research and Technology (QFort), National Cheng Kung University, Tainan 70101, Taiwan}

\author{Michael Barth}

\affiliation{Institute for Theoretical Physics, University of Regensburg, 93040 Regensburg, Germany}

\author{Aitor Garcia-Ruiz}

\affiliation{Department of Physics, National Cheng Kung University, Tainan 70101, Taiwan}

\affiliation{National Graphene Institute, University of Manchester, 
M13 9PL Manchester, U.K.}

\author{Alina Mre\'{n}ca-Kolasi\'{n}ska}

\affiliation{AGH University, Faculty of Physics and Applied Computer Science, 30-059 Krak\'{o}w, Poland}

\author{Ming-Hao Liu (劉明豪)}

\email{minghao.liu@phys.ncku.edu.tw}

\affiliation{Department of Physics, National Cheng Kung University, Tainan 70101, Taiwan}

\affiliation{Center of Quantum Frontiers of Research and Technology (QFort), National Cheng Kung University, Tainan 70101, Taiwan}

\author{Denis Kochan}

\email{denis.kochan@savba.sk}

\affiliation{Institute of Physics, Slovak Academy of Sciences, 84511 Bratislava, Slovakia}

\affiliation{Center of Quantum Frontiers of Research and Technology (QFort), National Cheng Kung University, Tainan 70101, Taiwan}

\begin{abstract}
Graphene-based van der Waals heterostructures take advantage of tailoring spin-orbit coupling (SOC) in the graphene layer by proximity effect.
At long-wavelength---saddled by the electronic states near the Dirac points---the proximitized features can be effectively modeled by the Hamiltonian involving novel SOC terms and allow for an admixture of the tangential and radial spin textures---by the so-called \emph{Rashba angle} $\theta_{\text{R}}$. Taking such effective models we perform realistic large-scale magneto-transport calculations---transverse magnetic focusing and Dyakonov-Perel spin relaxation---and show that there are unique qualitative and quantitative features allowing for an unbiased experimental disentanglement of the conventional Rashba SOC from its novel radial counterpart, called here the \emph{radial Rashba SOC}. Along with that, we propose a scheme for a direct estimation of the Rashba angle by exploring the magneto-response symmetries when swapping an in-plane magnetic field. To complete the story, we analyze the magneto-transport signatures in the presence of an emergent \emph{Dresselhaus SOC} and also provide some generic ramifications about possible scenarios of the \emph{radial superconducting diode effect}.
\end{abstract}

\maketitle
\end{CJK*}

Graphene-based van der Waals heterostructures provide an ideal platform for exploring various physical phenomena
\cite{HanKawakamiGmitraFabian_2014,
Roche_SpintronicsPerspective_2015,
Zutic2018,
Avsar2020:Colloquium,
Sierra_NatNanotech_2021}
that are intertwining single-particle and many-body properties, including magnetic, ferro-electric, topological and even superconducting states of matter 
\cite{
cao_unconventional_2018_9,
kennes_moire_2021_3,
jiang_spin_2022_6,
Szalowski_2023,
tschirhart_inspintorque_2023_7,
garcia-ruiz_mixed-stacking_2023,
tao_giant_2024_8}. A key ingredient for mediating many of these features is spin-orbit coupling (SOC). Although the latter is small in pristine graphene \cite{Gmitra2009,Konschuh2012:BLG}, it can be substantially enhanced by proximity effects of either adatoms or other layers 
\cite{weeks_QSH_2011_16,
gmitra_hydrogenated_2013_15,
Kaloni2014:APL,
wang_strong_2015,
Irmer2015-SOC-F-Graphene,
calleja_spatialvariation_2015_18,
gmitra_trivial_2016,
Alsharari2016PRB,
wang_origin_2016,
Zollner2016-SOC-Methyl,
Gmitra2017_Proceeding,
FrankPRB2017Copper,
kochan_model_2017,
volkl_magnetotrasport_2017,
Zollner2018,
zihlmann_spinrelaxation_2018,
Volkl2018,
ghiasi_chargespin_2019_14,
safeer_spinhalleffect_2020_23,
benitez_spingalvanic_2020_26,
ingla_electricalcontrol_2021_21,
tiwari_experimental_2022,
ontoso_unconventional_2023_13}. The strong SOC can be achieved by proximitizing graphene with transition metal dichalcogenides, a further interplay of twisting, stacking and of long supercell-periodicities allows one to engineer and tune the strengths of valley-Zeeman (VZ) and Rashba SOC 
\cite{gmitra_trivial_2016,
Gmitra2017_Proceeding,
cummings_lifetime_2017_31,
offidani_optimal_2017_32,
liyang_twistangle_2019_37,
david_heterobilayers_2019_39,
naimer_twistangle_2021_38,
vaneri_collinearEE_2022_40,
lee_twisted_2022_36,
peterfalvi_trilayers_2022_41}. 

The new kid on the block is a possibility to engender the Rashba SOC with the radial component. As shown in \cite{frank2024emergence}, there are very specific twist angles allowing for pure radial Rashba (RR) SOC. However, in general, the resulting spin texture admixes both \cite{Szalowski_2023,Zollner2023:PhysRevB.108.235166,yang2023twistangle}, the tangential and radial components, quantified by a Rashba angle $\theta_{\text{R}}$, and as such it gives rise to a tilted spin-momentum locking triggering, for example, unconventional spin-to-charge conversion \cite{lee_twisted_2022_36,vaneri_collinearEE_2022_40,Zollner2023:PhysRevB.108.235166}. These novel spin textures are going beyond the ordinary SOC-classification based on a reduction of just the graphene-point group symmetries~\cite{kochan_model_2017}. Indeed, the long-wavelength electronic states with momenta near the valley centers experience much more complex interaction landscape and the inherent supercell symmetries differ from an ordinary graphene allowing for more complex SOC Hamiltonians \cite{liyang_twistangle_2019_37,
david_heterobilayers_2019_39}.

\begin{figure}[b]
    \centering
    \includegraphics[width=\columnwidth]
    {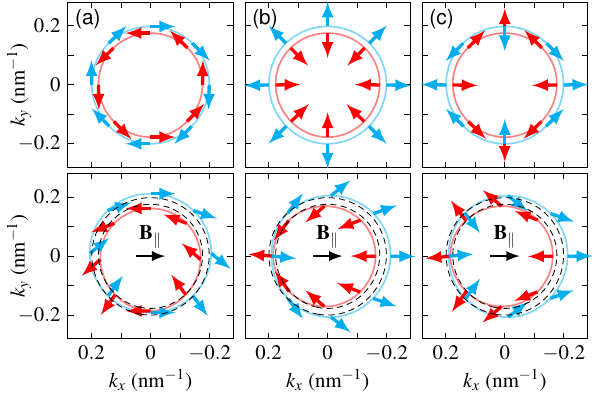}
    \caption{Fermi contours around the $K$ point at energy $E=-100$\,meV in the absence (upper panels) and presence (lower panels) of an in-plane magnetic field $\mathbf{B}_\parallel=(B_x,0,0)$ with $B_x=3$\,T. The directions how Fermi contours shift in the $k$-space corroborate with an increase or decrease of the spin-Zeeman energy, $\pm g\mu_\text{B} \langle s_x\rangle B_x$, as inherently imprinted by the underlying spin-textures: (a) CR, (b) RR and (c) Dresselhaus SOC, considering $\lambda_\text{CR}=\lambda_\text{RR}=\lambda_\text{D}=6\,$meV. The arrows on the contours display in-plane spin expectation values. The black dashed circles in the lower panels serve as guides to the eyes to contrast the original and shifted Fermi circles.}
    \label{fig:main_contour}
\end{figure}

\begin{figure*}[t]
    \centering
    \includegraphics[width=\textwidth]{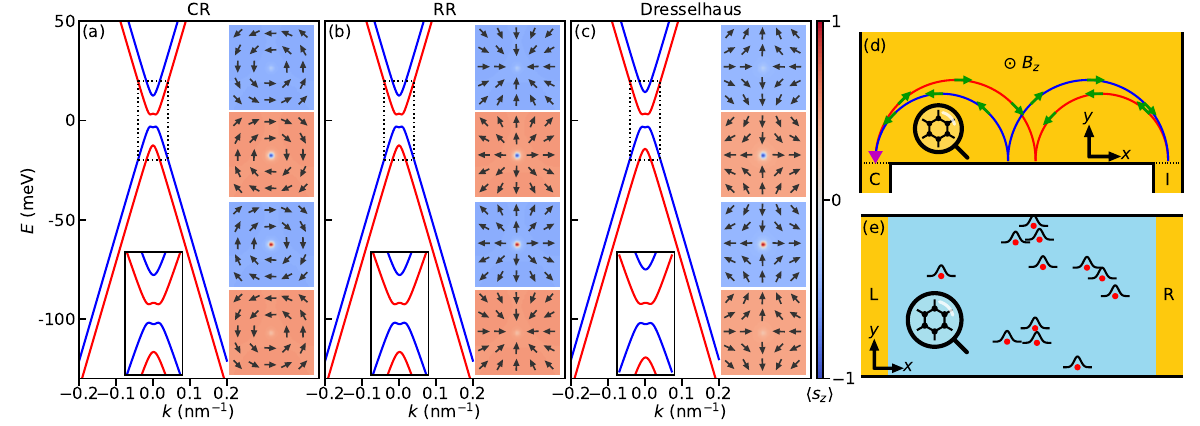}
    \caption{
    Calculated energy band structures and spin textures near the Dirac point of the proximitized graphene described by $\mathcal{H}_0+\mathcal{H}_\text{SOC}$, considering (a) $\mathcal{H}_\text{SOC}=\mathcal{H}_\text{VZ}+\mathcal{H}_\text{CR}$, (b) $\mathcal{H}_\text{SOC}=\mathcal{H}_\text{VZ}+\mathcal{H}_\text{RR}$, and (c) $\mathcal{H}_\text{SOC}=\mathcal{H}_\text{VZ}+\mathcal{H}_\text{D}$, in all three cases with $\lambda_{\text{VZ}}=3.5\,\text{meV}$ for $\mathcal{H}_\text{VZ}$, and separately with $\lambda_\text{CR}=\lambda_\text{RR}=\lambda_\text{D}=6\,\text{meV}$. The side panels at the 
    right of (a)--(c) show spin-textures---in-plane black arrows and red-blue color shading are displaying, respectively, $\langle s_x,s_y\rangle$ and $\langle s_z\rangle$, with the latter calibrated by the common color bar shown at the right of (c).
    Panels~(d)~and~(e)~display modeled devices: (d)~three terminal transverse magnetic focusing with schematics of spin-split skipping orbits, as well as, with the corresponding real-space spin-textures in the case of RR SOC---rotated by $90^\circ$ compared to $k$-space textures shown at (b), (e)~Dyakonov-Perel spin relaxation setup with schematics of Gaussian-type disorder.}
    \label{fig:main_band}
\end{figure*}

The main goal of this \emph{Letter} is to identify robust transport features allowing for experimentally unique discrimination between the conventional Rashba (CR) and RR SOC, and, moreover, also for the determination of the Rashba angle $\theta_\text{R}$. We show that SOC-imprinted spin textures---tangential and radial---typical for the CR and RR SOC---respond differently when exposed to an in-plane magnetic field $\mathbf{B}_\parallel=(B_x,B_y,0)$ rendering an additional spin-Zeeman interaction. Correspondingly, the initial spin-split Fermi contours shift---for CR SOC in perpendicular, and for RR SOC in parallel direction with respect to $\mathbf{B}_\parallel$; see Fig.~\ref{fig:main_contour} exemplified with $\mathbf{B}_\parallel=(B_x,0,0)$, to be further explained later.

Therefore, an in-plane magnetic field $\mathbf{B}_\parallel$ can resolve the contribution to a transport quantity $Q$ (probed along a direction given by unit vector $\hat{\mathbf{n}}$) coming from these two different spin-orbit textures.
In particular, $Q$ shows perfect swap-symmetry, $Q(\hat{\mathbf{n}}, \mathbf{B}_\parallel)=Q(\hat{\mathbf{n}}, -\mathbf{B}_\parallel)$ in two complementary cases: CR SOC when $\hat{\mathbf{n}}\parallel\mathbf{B}_\parallel$ and RR SOC when $\hat{\mathbf{n}}\perp\mathbf{B}_\parallel$,
which offer their unbiased experimental discrimination. Moreover, in the case when both CR and RR SOC admix via $\theta_\text{R}$ by changing an angle between $\hat{\mathbf{n}}$ and $\mathbf{B}_\parallel$ one finds a sweet-spot angle $\varphi=\sphericalangle (\hat{\mathbf{n}},\mathbf{B}_\parallel)$ at which $Q(\hat{\mathbf{n}}, \mathbf{B}_\parallel)=Q(\hat{\mathbf{n}}, -\mathbf{B}_\parallel)$. This sweet-spot angle is related to the Rashba angle $\theta_\text{R}=-\varphi$, up to modulus of adding $\pi$. The above symmetries persist even in the case the SOC Hamiltonian admixes a non-zero VZ SOC that offers unique experimental signatures for unveiling CR, RR and their admix angle $\theta_\text{R}$ while swapping the orientation and magnitude of the in-plane field $\mathbf{B}_\parallel$. 

Although our discussion and conclusions are general, for providing quantitative results we particularly focus on large-scale simulations \cite{liu_scalable_2015,rickhaus_snake_2015,mrenca-kolasinska_probing_2023,rao_ballistic_2023} of \emph{transverse magnetic focusing} (TMF) and the \emph{Dyakonov-Perel spin relaxation} (DPSR) \cite{Dyakonov1971,dyakonov_spin_1972,MeierZakharchenya:1984,Fabian:ActaPhysicaSlovaka:2007,boross_unified_2013}. For the former, $Q$ stands for the TMF-conductance $G$ (or more precisely its second peak, $G_\text{2nd}$), and a uniform out-of-plane magnetic field $\mathbf{B}_\perp = (0,0,B_z)$ is required. For the latter, no $\mathbf{B}_\perp$ is required, and $Q$ stands for the DPSR-rate, $\tau_s^{-1}$.
For completeness, we also consider, in addition to CR and RR, the Dresselhaus SOC \cite{Dresselhaus:PhyRev1958}, and reveal magnetotransport signatures of them. Even though there are not yet reported experimental, neither DFT evidences of such proximity-engineered term in graphene, the existing rush in the field of van der Waals materials can not preclude its future exposure.

\paragraph{Model Hamiltonian.} The effective tight-binding description of these spin features can be modeled by $\mathcal{H}=\mathcal{H}_0+\mathcal{H}_\text{Z}+\mathcal{H}_\text{SOC}$. Here, $\mathcal{H}_0$ represents the graphene Hamiltonian composed of conventional hopping $-t$ (taken as $-3$\,eV in the following) between all pairs of nearest-neighbor sites and
$\mathcal{H}_\text{Z}=g\mu_\text{B}\mathbf{s}\cdot\mathbf{B}$ stands for the spin-Zeeman coupling, where $\mu_\text{B}$ is the Bohr magneton, $\mathbf{s}=(s_x,s_y,s_z)$ is a vector of the $2\times 2$ Pauli matrices, $\mathbf{B}=\mathbf{B}_\parallel+\mathbf{B}_\perp=(B_x,B_y,B_z)$ is the net external magnetic field, and $g$ is an isotropic g-factor, we use $g=42.5$ as the elevated values of $g$ are typical for the confined graphene-based heterostructures with proximity engineered band splittings \cite{Frank:PhysRevB2020,Ge:NatMat2023,Xiang:NanoLetters2023}. The SOC Hamiltonian, $\mathcal{H}_\text{SOC}$, comprises 
VZ, CR, RR and Dresselhaus SOC terms, which can be respectively written as 
\begin{subequations}\label{eq Hsoc}
\begin{align}
    \mathcal{H}_{\text{VZ}} &= \frac{i}{3\sqrt{3}}\sum_{\langle\!\langle m, n\rangle\!\rangle}\sum_{\sigma}\lambda_{o(m)}c^{\dagger}_{n\sigma}\,\nu_{n\leftarrow m}\left[s_z\right]_{\sigma\sigma}\,c^{\phantom{\dagger}}_{m\sigma}\ ,\label{eq H_VZ}\\
    \mathcal{H}_{\text{CR}} &= \frac{2i}{3}\lambda_{\text{CR}}\sum_{\langle m, n\rangle}\sum_{\sigma,\sigma'}c^{\dagger}_{n\sigma}\,\left[\left(\mathbf{s}\times\hat{\mathbf{e}}_{n\leftarrow m}\right)\cdot\hat{\mathbf{e}}_z\right]_{\sigma\sigma'}\,c_{m\sigma'}\ ,\label{eq H_CR}\\
    \mathcal{H}_{\text{RR}} &=-\frac{2i}{3}\lambda_{\text{RR}}\sum_{\langle m, n\rangle}
    \sum_{\sigma,\sigma'}c^{\dagger}_{n\sigma}\,\left[\mathbf{s}\cdot\hat{\mathbf{e}}_{n\leftarrow m}\right]_{\sigma\sigma'}\,c_{m\sigma'}\ ,\label{eq H_RR}\\
    \mathcal{H}_{\text{D}} &= \frac{2i}{3}\lambda_{\text{D}}\sum_{\langle m, n\rangle}
    \sum_{\sigma,\sigma'}c^{\dagger}_{n\sigma}\,\left[\mathbf{s}^{*}\cdot\hat{\mathbf{e}}_{n\leftarrow m}\right]_{\sigma\sigma'}\,c_{m\sigma'}\ ,\label{eq H_D}
\end{align}
\end{subequations}
where $\sum_{\langle m, n\rangle}$ ($\sum_{\langle\!\langle m, n\rangle\!\rangle}$) denotes a summation over all pairs of lattice sites $m,n$ that are nearest (second-nearest) to each other, $c_{m\sigma}$ ($c^\dagger_{m\sigma}$) annihilates (creates) an electron with spin $\sigma$ on lattice site $m$, $\hat{\mathbf{e}}_{n\leftarrow m}$ appearing in \eqref{eq H_CR}--\eqref{eq H_D} is the unit vector pointing from site $m$ to site $n$, and $\left[\cdots\right]_{\sigma\sigma'}$ is the matrix element of $\left[\cdots\right]$ of the $\sigma$th row and $\sigma'$th column.

\begin{figure*}[t]
    \centering
    \includegraphics[width=\textwidth]{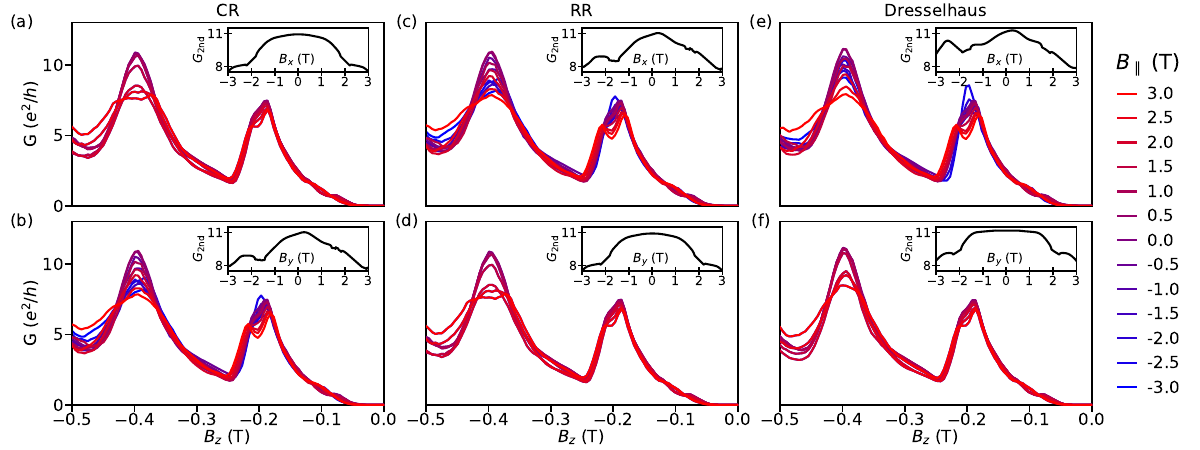}
    \caption{
    Simulated conductance, as a function of $B_z$ for
    a hole-doped graphene at energy $E=-100$\,meV, between the injector and detector (spaced by $1\,\mathrm{\mu m}$) of a three-terminal TMF device schematically shown in Fig.\ \ref{fig:main_band}(d), considering the CR~SOC [(a)~and~(b)], RR~SOC [(c)~and~(d)], and Dresselhaus~SOC [(e)~and~(f)], all subtended by a varying in-plane magnetic field of $\mathbf{B}_x$ for the upper panels and $\mathbf{B}_y$ for the lower panels. Color codes for different strengths of the in-plane magnetic field, $B_\parallel$, are shown at the right. The insets show the dependence of the second TMF conductance peaks, $G_\text{2nd}$, i.e., $G$ at $B_z \approx - 0.4$\,T, on the strength of the in-plane magnetic fields ($B_x$ for the upper panels and $B_y$ for the lower panels).}
    \label{fig:main_tmf}
\end{figure*}

In the VZ term, Eq.~\eqref{eq H_VZ}, $o(m)=A,B$ is the sublattice index of site $m$, and the sign factor $\nu_{n\leftarrow m}$ is $-1$ ($+1$) whenever the second-nearest hopping from site $m$ to site $n$ via their common nearest-neighbor forms a clockwise (counterclockwise) path. Note that the VZ Hamiltonian, Eq.~\eqref{eq H_VZ}, involves sublattice-resolved couplings, $\lambda_{A}$ and $\lambda_{B}=-\lambda_A$, such that $\lambda_\text{VZ}\equiv(\lambda_{B}-\lambda_{A} )/2$. The coupling constant $\lambda_\text{CR}$ in \eqref{eq H_CR}, where $\hat{\mathbf{e}}_z$ is the unit vector along the $z$ axis (and similarly $\hat{\mathbf{e}}_x,\hat{\mathbf{e}}_y$ to appear later), and $\lambda_{\text{RR}}$ in \eqref{eq H_RR} are parameterized by $\lambda_\text{CR}=\lambda_\text{R}\cos\theta_\text{R}$ and $\lambda_\text{RR}=\lambda_\text{R}\sin\theta_\text{R}$ via the Rashba coupling $\lambda_{\text{R}}>0$ and the Rashba angle $\theta_{\text{R}}$. In Eq.~\eqref{eq H_D}, $\lambda_\text{D}$ is the Dresselhaus SOC coupling strength, and $\mathbf{s}^*=(s_x,s_y,s_z)^*=(s_x,-s_y,s_z)$. Throughout the main text, we use $\lambda_\text{CR}=\lambda_\text{RR}=\lambda_\text{D}=6$\,meV, while $\lambda_\text{VZ}$ is set to be either $0$ or $3.5$\,meV. The low-energy band structure (centered at the $K$ point of the hexagonal Brillouin zone) of $\mathcal{H}_0+\mathcal{H}_\text{SOC}$ considering individually $\mathcal{H}_\text{SOC}=\mathcal{H}_\text{VZ}+\mathcal{H}_\text{CR}$, $\mathcal{H}_\text{SOC}=\mathcal{H}_\text{VZ}+\mathcal{H}_\text{RR}$, and $\mathcal{H}_\text{SOC}=\mathcal{H}_\text{VZ}+\mathcal{H}_\text{D}$ is shown in Fig.\ \ref{fig:main_band}(a), \ref{fig:main_band}(b), and \ref{fig:main_band}(c), respectively, along with the corresponding spin expectation values, $\langle\mathbf{s}\rangle=(\langle s_x\rangle,\langle s_y\rangle,\langle s_z\rangle)$, of the four bands, showing tangential, radial, and mixed in-plane spin textures. 

To confirm the expected symmetry $Q(\hat{\mathbf{n}}, \mathbf{B}_\parallel)=Q(\hat{\mathbf{n}}, -\mathbf{B}_\parallel)$ that corroborates the given type of Rashba/Dresselhaus SOC, we performed a series of bench marking large-scale magneto-transport simulations for TMF and DPSR schematically shown in Fig.\ \ref{fig:main_band}(d) and \ref{fig:main_band}(e), respectively, incorporating $\mathcal{H}_0+\mathcal{H}_\text{Z}+\mathcal{H}_\text{SOC}$ varying Fermi energy, SOC strengths, disorder, and $\mathbf{B}=\mathbf{B}_\parallel+\mathbf{B}_\perp$. Moreover, when employing the out-of-plane magnetic field $\mathbf{B}_\perp$ we also minimally couple the latter by means of the the Peierl's phase, as compared to the case with only $\mathbf{B}_\parallel$ that couples solely by means of the spin-Zeeman term $\mathcal{H}_\text{Z}$. 

\paragraph{Rationale.} Before showing the transport simulations, let us come back to Fig.\ \ref{fig:main_contour} (where $\lambda_\text{VZ}=0$) to further explain the shifts of the Fermi contours associated with an interplay of $\mathcal{H}_\text{SOC}$ and $\mathcal{H}_\text{Z}$ and its consequences for the swap symmetries. The upper row of Fig.\ \ref{fig:main_contour} shows Fermi contours at $E=-100$\,meV around the $K$ valley without $\mathbf{B}_\parallel$, along with the in-plane spin textures that are consistent with the lowest two bands of Figs.\ \ref{fig:main_band}(a)--(c).
With $\mathbf{B}_\parallel=(B_x,0,0)\equiv\mathbf{B}_x$ applied ($B_x=3\,$T), the lower panels of Fig.\ \ref{fig:main_contour} show distinct responses arising from the applied $\mathbf{B}_\parallel$ as expected from the characteristic SOC-imprinted spin-textures: Fermi surfaces in Fig.\ \ref{fig:main_contour}(a) for the CR SOC with the tangential textures move along $\pm\hat{\mathbf{e}}_y$, i.e.~perpendicular to $\mathbf{B}_x$, while these in Fig.\ \ref{fig:main_contour}(b) for the RR SOC with the radial textures, as well as in Fig.\ \ref{fig:main_contour}(c) for the Dresselhaus SOC, move along $\pm \hat{\mathbf{e}}_x$, i.e. parallel to $\mathbf{B}_x$. Changing the polarity from $\mathbf{B}_x$ to $-\mathbf{B}_x$ the contours will shift into the opposite sides (not shown). Importantly, for both $\pm \mathbf{B}_x$, the figures in Fig.\ \ref{fig:main_contour}(a) stay mirror-symmetric with respect to $\hat{\mathbf{e}}_y$, while those in Figs.\ \ref{fig:main_contour}(b) and \ref{fig:main_contour}(c) are mirror-symmetric with respect to $\hat{\mathbf{e}}_x$. Therefore, probing a transport quantity $Q$ along the $x$-axis for the in-plane field $\mathbf{B}_x$ will furnish fully symmetric magneto-responses, $Q(\hat{\mathbf{e}}_x,\mathbf{B}_x)=Q(-\hat{\mathbf{e}}_x,\mathbf{B}_x)$ and $Q(\hat{\mathbf{e}}_x,\mathbf{B}_x)=Q(\hat{\mathbf{e}}_x,-\mathbf{B}_x)$, for CR SOC, while \textit{a priori} asymmetric ones for RR and Dresselhaus SOC. Complementarily, probing $Q$ along the $y$-axis will give $Q(\hat{\mathbf{e}}_y,\mathbf{B}_x)=Q(-\hat{\mathbf{e}}_y,\mathbf{B}_x)$ and $Q(\hat{\mathbf{e}}_y,\mathbf{B}_x)=Q(\hat{\mathbf{e}}_y,-\mathbf{B}_x)$, for RR and Dresselhaus SOC, but not for CR SOC. The latter can be also recast in the alternative way, i.e.~as $Q(\hat{\mathbf{e}}_x,\mathbf{B}_y)=Q(-\hat{\mathbf{e}}_x,\mathbf{B}_y)$ and $Q(\hat{\mathbf{e}}_x,\mathbf{B}_y)=Q(\hat{\mathbf{e}}_x,-\mathbf{B}_y)$ where $\mathbf{B}_y\equiv(0,B_y,0)$ if one interchanges roles of the in-plane field and transport direction. 
This is a very robust and experimentally-sizable effect allowing to disentangle CR from RR SOC and even to determine $\theta_\text{R}$.

\begin{figure*}[t]
    \centering
    \includegraphics[width=\textwidth]{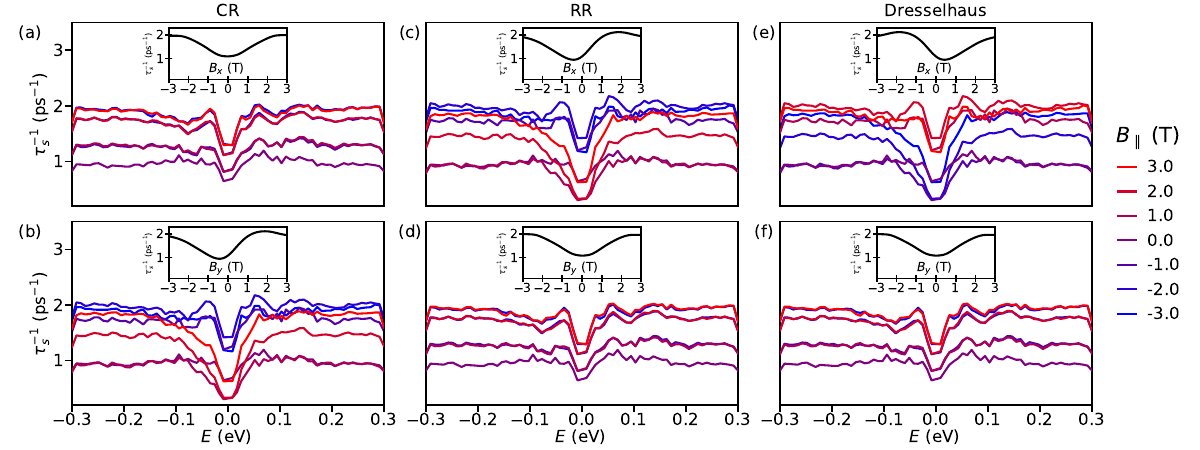}
    \caption{
    Simulated DPSR rate as a function of Fermi energy $E$ for a two-terminal disordered graphene schematically shown in Fig.~\ref{fig:main_band}(e), considering the CR [(a)~and~(b)], RR [(c)~and~(d)], and Dresselhaus~SOC [(e)~and~(f)], all subtended by a varying in-plane magnetic field of $\mathbf{B}_x$ for the upper panels and $\mathbf{B}_y$ for the lower panels. Color codes for different strengths of the in-plane magnetic field, $B_\parallel$, are shown at the right. The scattering region is a zigzag ribbon with the width $w=131a$ and length $L=300a$, where $a=2.46\,\AA$ is graphene lattice constant. 
    The insets show the variation of $\tau_s^{-1}$ with the in-plane magnetic fields ($B_x$ for the upper panels and $B_y$ for the lower panels) at $E=100$\,meV.}
    \label{fig:main_DPSR}
\end{figure*}

In our transport simulations, all based on the open-source \textsc{Python} package \textsc{Kwant}~\cite{groth_kwant_2014}, we explicitly verify these swap-symmetries  [panels (a), (d) and (f) in Figs.~\ref{fig:main_tmf} and 
\ref{fig:main_DPSR}] by taking $Q$ as the second conductance peak $G_{\text{2nd}}$ in the TMF simulation and the spin relaxation rate $\tau_s^{-1}$ in the DPSR simulation, considering separately CR, RR and Dresselhaus SOC with $\lambda_\text{CR}=\lambda_\text{RR}=\lambda_\text{D}=6\,\text{meV}$ and varying in-plane field components $B_x,B_y\in[-3,+3]$\,T.
Moreover, we confirm that in the case when CR and RR SOC terms admix with an angle $\theta_\text{R}$, one gets symmetry of $G_{\text{2nd}}(\hat{\mathbf{e}}_x,\mathbf{B}_\parallel (\hat{\mathbf{u}}))=G_{\text{2nd}}(\hat{\mathbf{e}}_x,-\mathbf{B}_\parallel (\hat{\mathbf{u}}))$ and 
$\tau_s^{-1}(\hat{\mathbf{e}}_x,\mathbf{B}_\parallel (\hat{\mathbf{u}}))=\tau_s^{-1}(\hat{\mathbf{e}}_x,-\mathbf{B}_\parallel (\hat{\mathbf{u}}))$ by probing a transport along $\hat{\mathbf{e}}_x$ while swapping the in-plane field $\mathbf{B}_\parallel (\hat{\mathbf{u}})$ along $\hat{\mathbf{u}}=\pm\left(\cos\theta_\text{R}, -\sin\theta_\text{R}, 0\right)$.
This is not hard to see when plotting the underlying Fermi contours and their spin textures using the SOC Hamiltonian consisting of both CR and RR terms. Generally, for a transport axis along $\hat{\mathbf{n}}$ 
one recovers symmetry $Q(\hat{\mathbf{n}}, \mathbf{B}_\parallel)=Q(\hat{\mathbf{n}}, -\mathbf{B}_\parallel)$
if $\varphi=\sphericalangle(\hat{\mathbf{n}},\mathbf{B}_\parallel)=-\theta_\text{R}+(\pi)$.
As $\mathcal{H}_\text{RR}$ and $\mathcal{H}_\text{D}$, are unitary equivalent, $s_x\mathcal{H}_\text{RR}s_x=\mathcal{H}_\text{D}$ assuming $\lambda_\text{RR}=-\lambda_\text{D}$ (or $\mathcal{H}_\text{RR}^{*}=\mathcal{H}^{\phantom{*}}_\text{D}$ assuming $\lambda_\text{RR}=\lambda_\text{D}$), TMF and DPSR cannot uniquely disentangle RR from the Dresselhaus SOC as both display the 
same swap-symmetry with $\mathbf{B}_\parallel$.



\paragraph{TMF.} Figure~\ref{fig:main_tmf} shows the conductance $G$ as a function of $B_z\in[-0.5,0]$\,T for a symmetric three-terminal TMF setup, schematically shown in Fig.~\ref{fig:main_band}(d), with
proximity-induced CR [panels~(a)~and~(b)], RR [panels~(c)~and~(d)] and Dresselhaus [panels~(e)~and~(f)] SOC.
Results are displayed for the hole-doped graphene at energy $E=-100$\,meV; for corresponding figures with $\lambda_{\text{VZ}}\neq 0$ and different values of $E$, $\lambda_{\text{CR}}$, $\lambda_{\text{RR}}$, and $\lambda_{\text{D}}$, see the Supplemental Material \footnote{See Supplemental Material for \ldots}. 
The insets in each panel show variations of the second conductance peak, $G_{\text{2nd}}(|\mathbf{B}_\parallel|)$, with either $\mathbf{B}_\parallel=\mathbf{B}_x$ or $\mathbf{B}_\parallel=\mathbf{B}_y$ \footnote{We focus on the second conductance TMF peak, because there two skipping orbits of the spin-split Fermi contour coalesce, see schematics in Fig.~\ref{fig:main_band}(d), what contrasts the first TMF peak that is becoming spin-split at elevated values of 
$\lambda_{\text{CR/RR/D}}$'s or at larger in-plane magnetic/Zeeman fields---see Fig.~\ref{fig:main_tmf} and also SM. Similarly, the higher TMF peaks are suffering 
from a more pronounced smearing---a finite width of Injector allows for an injection of electrons with varying incidence angles, while this variation grows with a propagation time, and with a bouncing of the skipping orbits off the edge the final signal gets more smeared.
It shall be emphasized that our results about the non-spin-split second TMF peak are valid assuming the Rashba and Dresselhaus SOC terms dominate over the in-plane Zeeman coupling, in the opposite case, the spin-textures of the skipping orbits and the Fermi contours subordinate to $\mathcal{H}_\text{Z}$.} displaying the expected swap-symmetry at panels (a),~(d)~and~(f), correlating the transport direction with the direction of $\mathbf{B}_\parallel$ and the given SOC.
In weaker in-plane fields ($|\mathbf{B}_\parallel| \lesssim 1\,\text{T}$), the $G_\text{2nd}$ shows no pronounced variations with $|\mathbf{B}_\parallel|$, but at higher fields
the variations 
can approach up to $30\%$. For more details, see the Supplemental Material \cite{Note1}.

\paragraph{DPSR.} Figure~\ref{fig:main_DPSR} shows the simulated DPSR-rates, $\tau_s^{-1}$, 
as functions of energy $E\in[-0.3,+0.3]$\,eV for a symmetric two-terminal setup, schematically shown in Fig.~\ref{fig:main_band}(e), considering CR [panels (a)~and~(b)], RR [panels (c)~and~(d)] and Dresselhaus SOC [panels (e)~and~(f)], in the presence of non-magnetic on-site disorder. We follow and extend a methodology developed in Refs.~\cite{Bundesmann:PRB2015,KochanBarth:PRL_2020,BarthKochan:PRB_2022} by incorporating an effect of 
the spin-Zeeman coupling, $\mathcal{H}_\text{Z}$, due to $\mathbf{B}_\parallel$. Electronic states in the scattering region are given in terms of the full model Hamiltonian, $\mathcal{H}_0+\mathcal{H}_\text{Z}+\mathcal{H}_\text{SOC}$, with twisted-boundary conditions across the zigzag edges~\cite{Bundesmann:PRB2015,KochanBarth:PRL_2020,BarthKochan:PRB_2022} to account for 2D transport features along $\hat{\mathbf{e}}_x$, while DPSR is due to a non-magnetic spatially-correlated Gaussian disorder $U(x,y)$ generated by means of the Thorsos-Monte-Carlo method~\cite{MackMethod_2013,Note1} \footnote{Disorder is defined by the on-site potential---on-site potential correlation function $C(|\mathbf{r}-\mathbf{r}'|)^\mathrm{2D} =\langle U(\mathbf{r}),U(\mathbf{r}')\rangle = K_0  \frac{(\hbar v_\mathrm{F})^2}{2\pi \xi^2} \exp\left(-\frac{|\mathbf{r}-\mathbf{r}'|^2}{2\xi^2}\right)$, as already used in other Dirac systems~\cite{BardarsonCorrelator:PRB2018}. The parameter $\xi$ is the spatial correlation length, $\xi=10a$, such that effective impurity concentration $\eta=\frac{\pi\xi^2}{wL}=0.799\%$. 
Furthermore, $K_0$ is a dimensionless scaling parameter which is set to 0.1, and we define Fermi velocity $v_\mathrm{F}$ such that $\hbar \tfrac{v_\mathrm{F}}{a}=\tfrac{\sqrt{3}}{2}t$ where $t=3$\,eV. Finally, we introduce periodic boundary conditions in a direction perpendicular to the transport axis, where we include a phase factor $\exp(i\phi)$ analogously to Refs.~\cite{Bundesmann:PRB2015,KochanBarth:PRL_2020,BarthKochan:PRB_2022}. The spin-flip relaxation rates are then obtained by averaging over 20 phases $\phi \in [0,2\pi]$ and 30 different disorder configurations.} with an effective impurity concentration $\eta=0.799\%$. Insets show $\tau_s^{-1}(\mathbf{B}_\parallel)$ at $E=100$\,meV, with expected swap-symmetry at panels (a), (d) and (f), correlating transport direction, with the direction of the in-plane field and with the type of SOC \footnote{As $\mathcal{H}_\text{RR}^{*}=\mathcal{H}^{\phantom{*}}_\text{D}$ Dresselhaus plots follow from RR ones by 
swapping the sign of $E$.}.

\paragraph{Radial supercurrent diode effect (SDE).} Finally, let us offer a short perspective how engendering exotic RR SOC in a superconducting two-dimensional electron gas (2DEG) would manifest in SDE. Considering quadratic 2DEG dispersion set by an effective mass $m^\ast$ and Fermi energy $E$, in the presence of RR SOC, $\alpha_\text{RR}(k_x s_{x} + k_y s_y)$, and an additional Zeeman-coupling, $\mathcal{H}_\text{Z}$, due to the in-plane field $\mathbf{B}_\parallel$, the Fermi contours acquire longitudinal shifts along $\mathbf{B}_\parallel$; see Fig.~\ref{fig:main_contour} with 2DEG Fermi contours around the $\Gamma$ point. Therefore, turning on the superconducting correlations will lead to an emergence of the helical phase, 
$\psi_\text{RR}(\mathbf{r})=e^{i \mathbf{P}_{\text{RR}}\cdot\mathbf{r}/\hbar}|\psi|$, with a finite center of mass momentum, $\mathbf{P}_{\text{RR}}\propto g \mu_\text{B} \alpha_\text{RR} \tfrac{m^\ast}{\hbar E} \mathbf{B}_\parallel$, aligned along $\mathbf{B}_\parallel$. This opposes the CR case \cite{Edelstein1989,Yuan2021,Kochan2023diode}, $\alpha_\text{CR}(k_x s_y - k_y s_x)$, where the center of mass momentum, $\mathbf{P}_{\text{CR}}$, develops in the direction perpendicular to $\mathbf{B}_\parallel$, i.e. 
$\psi_\text{CR}(\mathbf{r})=e^{i \mathbf{P}_{\text{CR}}\cdot\mathbf{r}/\hbar}|\psi|$, with $\mathbf{P}_{\text{CR}}\propto g \mu_\text{B} \alpha_\text{CR} \tfrac{m^\ast}{\hbar E} (\mathbf{B_\parallel}\times\hat{\mathbf{e}}_z)$. 
The immediate ramifications lead to different magneto-chiral responses
of the SDE critical current, $I^c_\text{SDE}$, and the Josephson inductance, $L_\text{J}$, on a probing current $\mathbf{j}$ and the in-plane field $\mathbf{B}_\parallel$. 
So both, $I^c_\text{SDE}=I^c_0 + \gamma_I f(\mathbf{j},\mathbf{B}_\parallel)$ and $L_\text{J}=L_{\text{J}0} + \gamma_L f(\mathbf{j},\mathbf{B}_\parallel)$, will depend on magneto-chiral factor $f(\mathbf{j},\mathbf{B}_\parallel)$ that, correspondingly, for RR or CR SOC equals to 
$\mathbf{j}\cdot\mathbf{B}_\parallel$ and $\mathbf{j}\cdot(\mathbf{B}_\parallel\times\hat{\mathbf{e}}_z)$, the magneto-chiral coefficients $\gamma_{I/L}$ encapsulate all material dependencies.
This unique magneto-chiral feature offered by RR SOC, if possible to realize in 2DEG, would allow to unveil 
the roots of SDE, namely, its SOC or screening-current origins.

\paragraph{Conclusions.} Although spectrally CR, RR and Dresselhaus SOC manifest similarly, the underlying wave functions and spin-textures discern when exposed to a Zeeman coupling triggered by 
an in-plane magnetic field. We showed that such SOC-imprinted features allow to disentangle CR and RR SOC components, quantified by the Rashba angle $\theta_\text{R}$, when exploring the magneto-transport symmetries of TMF and DPSR signals by swapping the orientation of in-plane field. While the simulated TMF provides different quantitative responses for all three SOC cases allowing for their mutual discrimination, the DPSR lacks this ability in the case of RR and Dresselhaus SOC, however it allows to discriminate the latter from the CR SOC. Even though we focused on graphene with the proximity engendered SOC our conclusions are valid in general and offer novel magneto-chiral response of the SDE.

\begin{acknowledgments}
We thank Jaroslav Fabian and Paulo~E.~Faria Junior for useful discussions. W.-H.K., A.G.-R., and M.-H.L.\ gratefully acknowledge National Science and Technology Council of Taiwan (grant no.\ NSTC 112-2112-M-006-019-MY3) for financial support and National Center for High-performance Computing (NCHC) for providing computational and storage resources. M.B.~acknowledges support by the Deutsche Forschungsgemeinschaft (DFG, German Research Foundation) within Project-ID 314695032--SFB 1277 (project A07). A.M.-K.~acknowledges partial support by program ``Excellence initiative - research university'' for the AGH University of Krakow, and by PL-Grid Infrastructure. D.K.~acknowledges partial support from the project IM-2021-26 (SUPERSPIN) funded by the Slovak Academy of Sciences via the programme IMPULZ 2021, and Grant DESCOM VEGA 2/0183/21.
A.G.-R., A.M.-K., M.-H.L. and D.K. acknowledge the hospitality of Ming-Yu Liu for organizing a trip to Fo Guang Shan Monastery issuing in many fruitful discussions. 
\end{acknowledgments}

\bibliographystyle{apsrev4-2}
\bibliography{reference}

\end{document}